\documentclass[]{aa}
 
\usepackage{multirow}
\usepackage{multicol}

\usepackage{graphicx}
\usepackage{txfonts}

\usepackage{color}
\usepackage[colorlinks=true,allcolors=blue]{hyperref}
\usepackage{natbib}
\bibpunct{(}{)}{;}{a}{}{,} 
\usepackage{paralist}
\usepackage{epstopdf}
\usepackage{epsfig}
\usepackage{amsmath}
\usepackage{tablefootnote}
\usepackage{amssymb}
\usepackage{bm}
\usepackage{tabularx}
\usepackage{lmodern}
\usepackage{lineno}
\renewcommand{\linenumbers}{}

\begin{document} 

    \title{A deep {\it HST} view of the open cluster NGC\,2158: binaries, mass functions, and M-dwarf discontinuity} 
       \author{
       A.\,V.\,Marchuk$^{1}$\thanks{E-mail: anastasiavadimovna.marchuk@studenti.unipd.it},
        F.\,Muratore$^{1}$,
        A.\,P.\,Milone$^{1,2}$, 
        M.\,V.\,Legnardi$^{1}$,
        F.\,D'Antona$^{3}$,
        G.\,Cordoni$^{4}$,
       A.\,Mastrobuono-Battisti$^{1,2,5}$,    
        E.\,Bortolan$^{1}$,
        F.\,Dell'Agli$^{3}$,
        E.\,Dondoglio$^{2}$,
        E.\,P.\,Lagioia$^{6}$,
        A.\,F.\,Marino$^{2}$, 
        M.\,Tailo$^{7}$,
        C.\,Ventura$^{3}$,
        P.\,Ventura$^{3}$,
        T.\,Ziliotto$^{1}$
        }

\institute{$^1$  Dipartimento di Fisica e Astronomia ``Galileo Galilei'', Universit\'a di Padova, Vicolo dell'Osservatorio 3, 35122 Padova, Italy \\
$^{2}$ Istituto Nazionale di Astrofisica - Osservatorio Astronomico di Padova, Vicolo dell'Osservatorio 5, 35122 Padova, Italy\\
$^{3}$ Istituto Nazionale di Astrofisica, Osservatorio Astronomico di Roma, Via Frascati 33, 00077 Monte Porzio Catone, Italy\\
$^{4}$ Research School of Astronomy \& Astrophysics, Australian National University, Canberra, ACT 2611, Australia \\
$^{5}$  Dipartimento di Tecnica e Gestione dei Sistemi Industriali, Universit\'a degli Studi di Padova, Stradella S. Nicola 3, I-36100 Vicenza, Italy\\
$^{6}$
South-Western Institute for Astronomy Research, Yunnan University, Kunming, 650500 P. R. China \\
$^{7}$  Dipartimento di Fisica e Astronomia ``Augusto Righi'', Universit\'a di Bologna, Via Gobetti 93/2, 40129 Bologna, Italy
}

\titlerunning{A deep HST view of NGC\,2158.} 
\authorrunning{Marchuk et al.}

\abstract{
A significant fraction of stars in both the Galactic field and stellar clusters belong to binary systems. Understanding their properties is therefore fundamental for a comprehensive picture of stellar structure, stellar evolution, and cluster dynamics. Despite extensive work on binaries in clusters, key questions remain open, particularly concerning photometric binaries among low-mass stars. While the binary fraction among field stars shows a strong dependence on stellar mass, studies of star clusters have so far suggested an approximately constant fraction across the limited mass range explored. Moreover, the mass function (MF) of very low-mass stars remains poorly constrained in clusters older than a few hundred Myr.

In this work, we use deep \textit{Hubble Space Telescope} imaging of the intermediate-age open cluster NGC\,2158 to investigate its binary population and derive the luminosity and MFs down to $\sim0.14\,M_{\odot}$. This dataset enables the first detailed study of binaries in this cluster.

We measure a global binary fraction of $38\%$, consistent with that observed in other open clusters, and detect a clear mass dependence: the fraction decreases from $\sim52\%$ at $1.0 M_{\odot}$ to $\sim11\%$ at $0.2 M_{\odot}$. This trend mirrors that seen in the Galactic field, suggesting that binaries in NGC\,2158 and field populations share similar properties. The MF of NGC\,2158 is best described by three regimes: high-mass stars ($\alpha = -2.49 \pm 0.19$), low-mass stars ($\alpha = -1.11 \pm0.09$), and very low-mass stars ($\alpha = -0.08\pm0.07$). The slope change near $1.0\,M_{\odot}$ agrees with recent open cluster surveys, although we find a deficit of stars at the lowest masses ($M \lesssim 0.3\,M_{\odot}$).

Finally, we identify a discontinuity in the main sequence around $M \sim 0.3\,M_{\odot}$. We explore the possibility that this feature traces the $^3$He-driven instability predicted by stellar models, analogous to the “Jao Gap” observed in the color–magnitude diagram of nearby field stars.

}
 

\maketitle

\section {Introduction}
\label{sec:intro}

Binary stars are key members of stellar populations and are central to many aspects of cluster astrophysics. The evolutionary paths of binary systems can differ significantly from those of single stars, depending on their orbital properties and mass ratios, and they are progenitors of exotic stellar populations including blue stragglers, cataclysmic variables, and millisecond pulsars. Accurate determinations of global parameters such as the mass function (MF) and the total mass of a stellar population require precise knowledge of the frequency and the properties of binary stars.
\begin{table*}\label{tab:data}
\caption{Information on the images used in this work. All observations were obtained with the WFC/ACS on board the {\it HST}, as part of program GO\,10500 (PI: L.\,Bedin).}    \centering
    \begin{tabular}{lcc}
    \\
 Date & N $\times$ exposure time& Filter    \\
    \hline
2005 Oct 24-25  & 2$\times$20s$+$1145s$+$4$\times$1150s    &  F606W \\
2005 Oct 24-25  & 1s$+$2$\times$20s$+$1130s$+$4$\times$1150s    &  F814W \\    
2006 Nov 06  & 1s$+$4$\times$20s$+$1130s$+$1145s$+$8$\times$1150s    &  F606W \\
  \hline
    \end{tabular}
    \label{dataset_table}
\end{table*}
Binary frequency and properties influence both the structural evolution of clusters and the interpretation of their observable features. They must be accounted for in order to fully understand the dynamical state of a cluster. Through their role as internal energy sources, binaries regulate the long-term dynamical evolution of clusters.

The color–magnitude diagram (CMD) provides one of the most effective means to investigate binary populations in resolved stellar populations. Because unresolved binaries appear offset from the main sequence (MS), CMD analyses allow statistical determinations of binary fractions with no dependence on orbital inclination or period. This approach has been successfully applied to numerous star clusters \citep[e.g.,][]{romani1991a, bolte1992a, cool2002a, bellazzini2002a} and in nearby dwarf galaxies \citep{legnardi2025a, muratore2025a}.

Accurate photometry from high-resolution Hubble Space Telescope (HST) images and from recent Gaia data releases has enabled systematic investigations of large samples of open and globular clusters. Among the main achievements, these studies have revealed that the fraction of binary stars varies significantly from one cluster to another \citep[][]{sollima2007a, milone2012a, milone2016a, cordoni2023a, mohandasan2024a}. In particular, the binary fraction in cluster cores correlates with cluster mass: massive globular clusters often exhibit very small core binary fractions, as low as 2–5\%, whereas low-mass open clusters exhibit core binary fractions up to an order of magnitude higher \citep[e.g.,][and references therein]{mohandasan2024a}. However, a more controversial result that has emerged from studies of binaries in star clusters typically shows a nearly constant binary fraction as a function of the primary stellar mass. This finding appears to contrast with observations in the Galactic field, where a strong correlation between binary fraction and primary mass is well established. In the field, stars with masses above $\sim$10 M$_{\odot}$ are predominantly found in binaries, whereas the binary fraction decreases to below 20\% for stars with masses smaller than $\sim$0.2 M$_{\odot}$ \citep[][and references therein]{offner2023a}. Since a large fraction of field stars originated in clusters that have since dissolved, this discrepancy between cluster and field observations calls for further investigation.

Beyond binaries, the stellar MF provides a fundamental diagnostic of cluster formation and dynamical evolution. The canonical \citet{kroupa2001a} MF has long been considered a universal description of stellar populations, yet recent evidence from the survey of open clusters based on Gaia DR3 data by \citet{cordoni2023a} suggests significant deviations from this form in the analyzed mass interval ($\sim 0.3$–2.5 M$_{\odot}$). Extending MF determinations to lower stellar masses with deep HST data provides a crucial opportunity to test these deviations, and to assess the role of dynamical evolution in shaping the MF.

At the faintest end of the MS, the so-called M-dwarf gap has recently been identified in field populations \citep[e.g.,][]{jao2018a}, where it manifests as a discontinuity in the number counts of late-type stars. This feature is thought to arise from changes in stellar interior structure at the transition to full convection. Extensive photometric surveys with HST and James Webb Space Telescope have been conducted, mostly focused on globular clusters \citep[e.g.,][]{richer2006a, dondoglio2022a, marino2024a}. However, no such gap has yet been observed, although the homogeneous stellar populations of clusters provide an ideal laboratory for testing its presence and properties. Detecting (or ruling out) gaps or discontinuities along the M-dwarf sequence in a cluster environment would thus represent a significant step forward in connecting stellar astrophysics with cluster population studies.

In this paper, we address these three issues by analyzing deep HST photometry of the $\sim$2-Gyr-old Galactic open cluster NGC\,2158 \citep{carraro2002a, bedin2010a}. We investigate its CMD across a wide range of stellar masses to study the binary population, derive the cluster MF, and search for evidence of the M-dwarf gap or discontinuity. The paper is organized as follows: in Section\,\ref{sec:data}, we describe the dataset and the procedures used to derive stellar astrometry and photometry. Sections\,\ref{sec:binaries} and \ref{sec:MF} present the analysis of the binary population and the stellar MF, respectively. In Section\,\ref{sec:gap}, we focus on the faint end of the MS and on the discontinuity detected along the M-dwarf sequence. Finally, Section\,\ref{sec:summary} provides a summary of our results and a discussion of their implications.

\begin{figure*}
    \centering
    \includegraphics[width=1.0\linewidth]{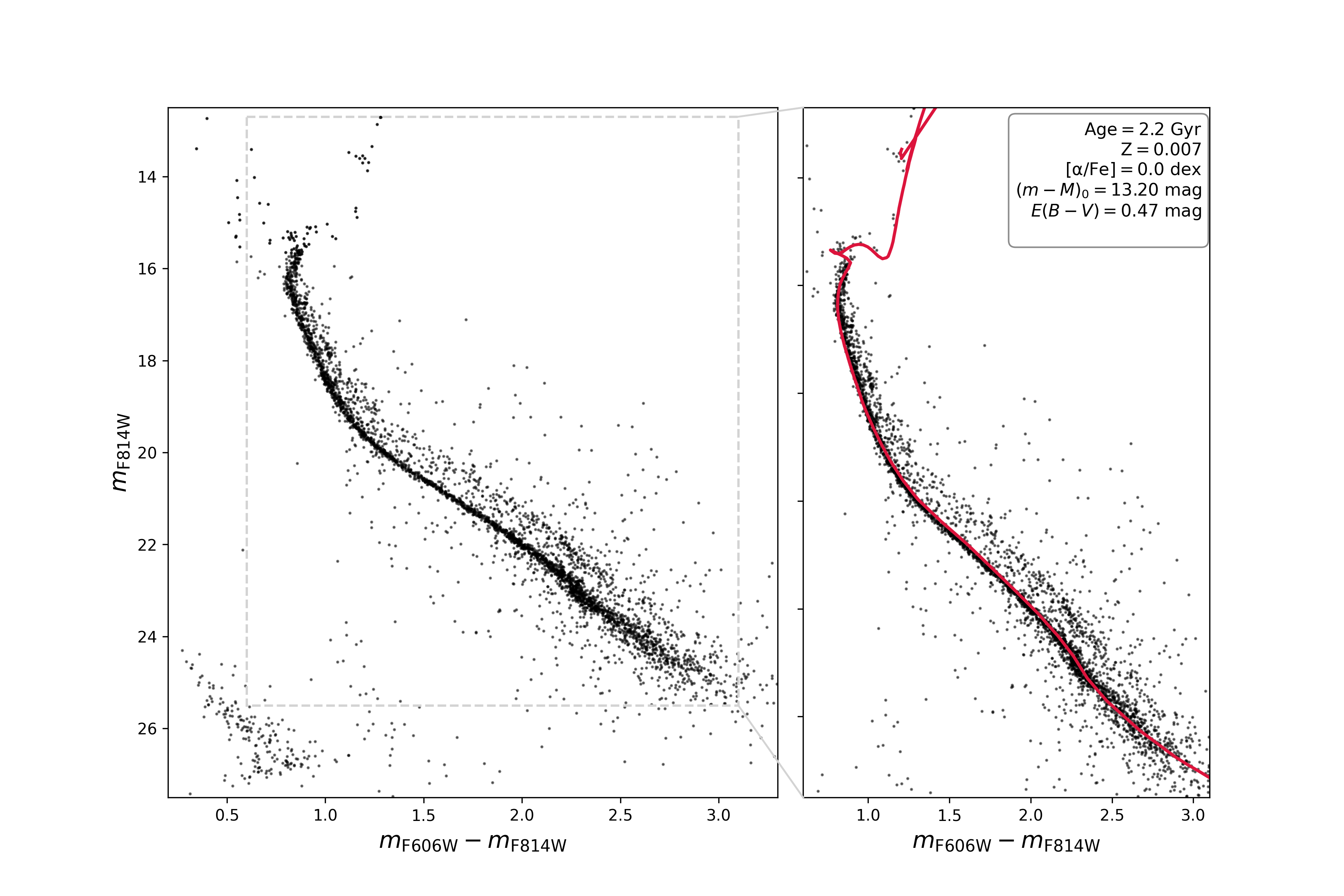}
\caption{$m_{\rm F814W}$ vs.\,$m_{\rm F606W}-m_{\rm F814W}$ CMD of NGC\,2158 corrected for differential reddening (left panel). The right panel shows a zoomed-in view of the same CMD, where the best-fit isochrone from \citet[][]{marigo2017a} is overplotted.}
    \label{fig:cmd}
\end{figure*}
\section{Data and data reduction}\label{sec:data}
To investigate NGC\,2158, we used images collected with the Wide Field Channel of the Advanced Camera for Surveys (WFC/ACS) on board the {\it HST}, whose main characteristics are summarized in Table\,\ref{dataset_table}.
We performed all measurements on the $\_$flc images, which are bias-, dark-, and flat-field–corrected exposures that include the pixel-based correction for charge transfer efficiency \citep[CTE,][]{anderson2010a}. 

We measured ACS/WFC photometry using the Jay Anderson's KS2 reduction pipeline \citep[an advanced descendant of the \texttt{kitchen\_sync} package, ][]{anderson2008a}, and employing precisely Method 1 and Method 2 \citep{sabbi2016a, bellini2017a, milone2023a}. The two methods differ in the way fluxes and positions are extracted from the images:

\begin{description}
  \item[Method 1:] For each exposure, we model and subtract neighboring stars, then perform a point-spread function (PSF) fit using a spatially variable effective PSF (ePSF) model \citep{anderson2000a}. This yields independent estimates of the flux and position per star per exposure. The final photometry and astrometry are computed by averaging across all exposures, taking into account measurement uncertainties and systematic offsets between frames.

  \item[Method 2:] After neighbor subtraction, we perform aperture photometry on the residual images by summing counts within a small fixed aperture around each star. This method is particularly useful for fainter stars whose profile is not well captured by a PSF fit. The aperture fluxes are then combined over multiple exposures.
\end{description}

We adopt Method 1 for bright stars, where the PSF model is well constrained, and Method 2 for the measurement of fainter stars or in crowded regions. All measurements are corrected for geometric distortion, flat-field effects, and charge-transfer inefficiency according to the standard ACS/WFC calibration recipes. Cross-epoch transformations and frame-to-frame photometric offsets are derived from bright, well-measured stars. The final catalog lists for each star the best flux, magnitude, uncertainty, and positional information, derived via the combination of the two methods.
Stellar positions were corrected for geometric distortion using the solution provided by \citet{anderson2022a}, and the photometry, calibrated to the VEGA-mag system as in \citet{milone2023a}, was corrected for the effects of differential reddening following the procedures described in \citet{milone2012a} and \citet{legnardi2023a}.

The resulting $m_{\rm F814W}$ vs. $m_{\rm F606W}-m_{\rm F814W}$ CMD is shown in the left panel of Fig.\,\ref{fig:cmd}. A visual inspection of this CMD reveals a well-defined MS, extending over more than nine magnitudes in the F814W band. A prominent binary sequence, composed of MS–MS systems, runs parallel to the MS on its red side. The upper MS exhibits a sudden change in slope around $m_{\rm F814W} \sim 16.3$, which we tentatively associate with the gaps predicted by theoretical studies \citep{bohmvitense1970a, dantona2002a} and observed in nearby Galactic open clusters \citep{bohmvitense1974a, debruijne2000a, debruijne2001a} and intermediate-age Magellanic Cloud clusters \citep{milone2023a}. The white-dwarf cooling sequence is clearly visible in the lower-left region of the CMD \citep[see][for a detailed study of the white-dwarf sequence in NGC\,2158]{bedin2010a}. The right panel of Fig.\,\ref{fig:cmd} also shows the best-fit isochrone from \cite[][]{marigo2017a}, corresponding to a stellar population with an age of 2.2\,Gyr, metallicity $Z=0.007$, and [$\alpha$/Fe]\,=\,0.0. We adopted a distance modulus of $(m-M)_0=13.20$\,mag and a foreground reddening of $E(B-V)=0.47$\,mag.

To characterize photometric uncertainties, completeness levels, and blending effects in the CMD, we performed extensive artificial-star (AS) tests following the procedures described by \citet{milone2012a,milone2023a}. We generated a large catalog of one million ASs with the same radial distribution and luminosity-function shape as the observed stars. Each AS was assigned a magnitude and color according to the empirical MS fiducial, covering the full observed range. The stars were added one at a time (so that no two ASs overlap). Then, the full photometric reduction was repeated, including ePSF fitting and the same selection criteria adopted for real stars. An AS was considered recovered if its output magnitude differed from the input value by less than 0.75 mag and its measured position was within $\sim$0.5 pixel of the input coordinates. The ratio of recovered to injected ASs, computed in bins of magnitude and radial distance, provided completeness fractions as a function of magnitude and position. Photometric uncertainties, evaluated in magnitude bins, were derived from the dispersion of the (output–input) magnitude distributions for recovered artificial stars.

\section{The population of binaries}\label{sec:binaries}

At the distance of NGC\,2158, most of the binary systems that are able to survive in the dense cluster environment are not spatially resolved. These systems therefore appear as point-like sources whose observed flux is the sum of the light from both components. In the following, we focus on the population of unresolved binaries, while Section\,\ref{subsec:wide} is devoted to wide binaries that may be resolved by {\it HST}
To estimate the fraction of  unresolved binaries with mass ratio $q>0.5$, we adopted the procedure described by \citet{milone2012a}, summarized as follows. 

\begin{figure}
    \centering
    \includegraphics[width=1.0\linewidth]{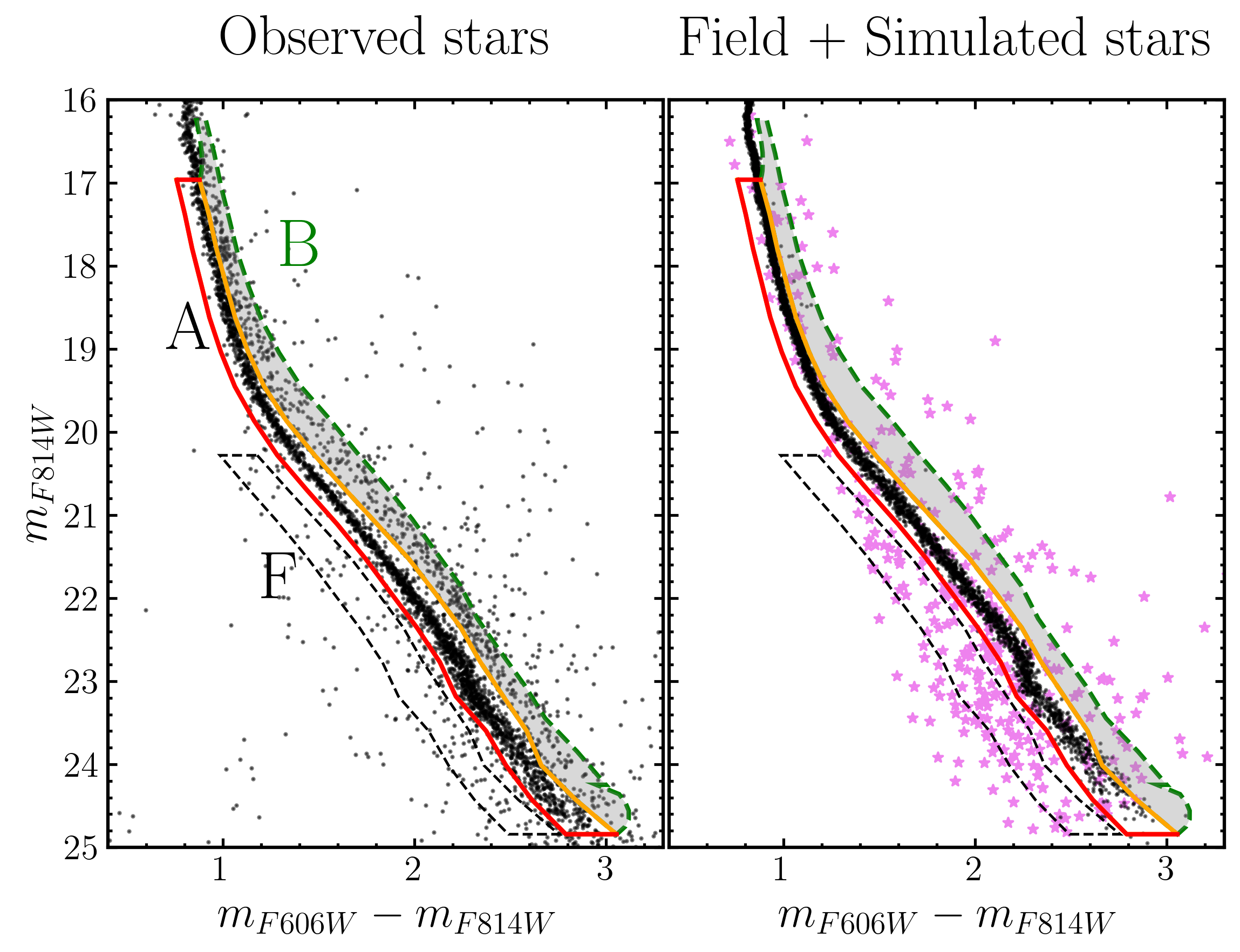}
    \caption{{\it Left}: Zoom-in of the left-panel CMD shown in Figure\,\ref{fig:cmd} around the region where binaries with large mass ratios can be more clearly distinguished from the bulk of single stars. {\it Right}: Simulated CMD including only single stars. Simulated cluster members and field stars are shown as black points and magenta starred symbols, respectively. The red and green lines in both panels enclose the region A of the CMD, used to infer the binary fraction. The gray shaded area delineates region B, a subregion of A that primarily hosts binaries with mass ratios larger than 0.5. Region F, populated by field stars, is the CMD area enclosed by the black dashed lines. }
    \label{fig:binaries}
\end{figure}

Two regions were defined in the CMD. Region \textit{A} includes single MS stars and binaries with primary magnitudes $17.0 < m_{\rm F814W} < 24.8$ (area enclosed by red and green lines in Fig.~\ref{fig:binaries}). Region \textit{B}, a subset of \textit{A}, contains binaries with $q>0.5$ (shaded area in Fig.~\ref{fig:binaries}). The green dashed line represents the locus of equal-mass binaries shifted by $+4\sigma$ in color, where $\sigma$ is the photometric uncertainty following \cite{milone2012a}. The red line marks the MS fiducial, offset by $-4\sigma$. The binary locus for mass ratio, $q=0.5$ (orange line in Fig.~\ref{fig:binaries}) was derived from the mass-luminosity relation of the best-fitting \cite{marigo2017a} isochrones. 

The binary fraction was computed as in Eq.~1 of \cite{milone2012a}:
\begin{equation}
f_{\rm bin}^{q>0.5} = 
\frac{N_{\rm REAL}^{B} - N_{\rm FIELD}^{B}}
     {N_{\rm REAL}^{A} - N_{\rm FIELD}^{A}}
     - \frac{N_{\rm AS}^{B}}{N_{\rm AS}^{A}} ,
\end{equation}
where $N_{\rm REAL}^{A,B}$ are the completeness-corrected numbers of cluster stars in regions \textit{A} and \textit{B}, and $N_{\rm FIELD}^{A,B}$ and $N_{\rm AS}^{A,B}$ are the corresponding numbers of field stars normalized to the cluster field area and of ASs.

To derive the simulated CMD used to estimate $N_{\rm AS}^{\rm A,(B)}$, we adopted the same procedure described by \citet{milone2012a}. This ensures that the AS sample used to infer the fraction of chance-superposition binaries, i.e.\ the ratio $\frac{N_{\rm AS}^{\rm B}}{N_{\rm AS}^{\rm A}}$, reproduces the spatial and luminosity distributions of the observed stars.

The number of field stars was estimated from a synthetic $m_{\rm F814W}$ versus $m_{\rm F606W}-m_{\rm F814W}$ CMD, generated with the \texttt{Trilegal} code\footnote{\url{http://stev.oapd.inaf.it/cgi-bin/trilegal}} \citep{girardi2005a}. The simulations provide the expected star counts, magnitudes, and colors in the field of view of NGC\,2158 based on the Galactic model of \citet{girardi2005a}. To reduce uncertainties in the estimated number of field stars arising from possible discrepancies between the Galactic model and the observed field, we identified a region F in the CMD that is expected to be populated  mainly by field stars. 
 To account for binary systems composed of MS stars and WDs that may populate region F of the CMD, we adopted the following procedure. We started simulating 183 WDs, as inferred by \cite{bedin2010a} from the same dataset used in this study, and assumed that 40\% of them are in MS–WD binary systems. For these binaries, we added a MS companion to the WD flux, with the companion’s luminosity drawn from the observed luminosity function of MS stars. Based on this analysis, we expect that less than ten binary systems can populate region\,F.
We then scaled the number of simulated field stars such that the total number of simulated stars in region F is ten times larger than the number of observed stars  after accounting for the contamination of WD-MS binaries of NGC\,2158. This factor of ten ensures that statistical fluctuations are minimized and that the comparison is not dominated by low-number statistics. 

We find a binary fraction of $f_{\rm bin}^{q>0.5}=0.19 \pm 0.01$ for systems with mass ratios greater than 0.5. By assuming a flat mass-ratio distribution and adopting a minimum secondary mass of 0.075 M$_{\odot}$ \citep{offner2023a}, we extrapolate a total binary fraction of $f_{\rm bin}=0.38 \pm 0.02$. 
 To estimate the uncertainties on the binary fraction, we simulated a CMD containing 10$^{6}$ stars, adopting the observed binary fractions. From this simulated CMD, we randomly extracted the same number of stars as in the observed CMD and measured the binary fraction using the same procedure applied to the real data. This process was repeated 1,000 times, and the uncertainty on the binary fraction was taken as the root-mean-square (rms) of the 1,000 measurements.
 These values are comparable to those observed in many open clusters \citep{cordoni2023a}.

\subsection{Wide binaries}
\label{subsec:wide}
 Given WFC/ACS's high angular resolution and the distance of the cluster, a significant fraction of wide binaries can be resolved into two distinct sources. For example, adopting the log-normal distribution for the orbital separations of binary systems derived by \citet{raghavan2010a} for field stars and assuming an average stellar mass of $0.6\,M_{\odot}$, we would expect that approximately 28\% of all binaries have projected separations larger than 220~AU, while about 5\% have separations exceeding 8,000~AU.

To assess the ability of our procedure to resolve and accurately measure the individual components of binary systems, we performed extensive AS tests. We generated a synthetic catalog composed exclusively of binary systems, adopting a flat mass-ratio distribution and random separations uniformly distributed between 0 and 36 ACS/WFC pixels, corresponding to $0$-$8,000$~AU at the distance of NGC\,2158. The primary components were drawn to follow the same radial distribution and luminosity function as the observed stars. These ASs were then processed using exactly the same reduction and analysis applied to the real data.

Based on the analysis of binaries with mass ratio larger than 0.5, we find that binary systems with separations smaller than 1.0 ACS/WFC pixel (i.e., $\sim220$~AU) are not resolved and appear as point-like sources, with a measured flux consistent with the sum of the fluxes of the two components. In contrast, systems with separations larger than 2 ACS/WFC pixels (i.e., $\sim440$~AU) are fully resolved. For intermediate separations, both resolved and unresolved binaries are detected, with systems having larger mass ratios being resolved at smaller separations than those with low mass ratios.

To estimate the fraction of resolved binaries, we adopted the following procedure. 
We first determined the observed number of apparent double MS systems, 
$N_{\rm REAL}^{\rm double}$, defined as MS stars that have a resolved, fainter MS companion within 36 pixels and with a mass differing more than 0.5 times the mass of the primary star.

In an analogous way, we computed the number of apparent double ASs, $N_{\rm AS}^{double}$, 
that is, the number of resolved ASs with an observed MS neighbour having a mass ratio larger than 0.5.

We defined as probable MS members all stars lying between the red and yellow lines in 
Fig.\,\ref{fig:binaries} and only considered stars with $17.0 < m_{\rm F814W} < 23.0$ mag.

To account for contamination from chance superpositions with field stars, we estimated the number of double field stars, $N_{\rm FIELD}^{double}$.
We assigned random coordinates to the field population and identified those field stars falling in 
the MS region of the CMD that have an observed MS star within 36 pixels. For each such pair, 
we computed the mass of a cluster star with the same magnitude as the field star and 
classified the system as a double field star when the mass ratio between the fainter and the brighter component exceeds 0.5.

The fraction of resolved binaries with mass ratio larger than 0.5 is calculated as:
\begin{equation}
f_{\rm resolved\,bin}^{q>0.5} = 
\frac{N_{\rm REAL}^{double} - N_{\rm FIELD}^{double}}
     {N_{\rm REAL}^{A} - N_{\rm FIELD}^{A}}
     - \frac{N_{\rm AS}^{double}}{N_{\rm AS}^{A}} ,
\end{equation}

We find a negligible fraction of wide binaries, with 
$f_{\rm resolved\,bin}^{q>0.5} = 0.01 \pm 0.01$. 
Restricting the analysis to systems with projected separations smaller than 1000\,AU leads to the same conclusion: 
the fraction of wide binaries in NGC\,2158 is consistent with zero, in sharp contrast to what is observed among field stars. As a consequence, in the following, we consider results from unresolved binaries only.

\subsection{The binary fraction as a function of primary-star luminosity and mass}

We investigated how the incidence of binaries depends on the brightness and mass of the primary component. The fraction of systems with $q > 0.5$ was measured in five 1.6 mag-wide bins along the F814W MS, following the same procedure described above. The results are provided in Table\,\ref{tab:binaries}, where we also provide the minimum and maximum mass of single stars and primary components of the binary systems in each bin, along with the total fraction of binaries. The total  fraction of binaries is extrapolated from the measured fraction of binaries with q$>$0.5 by assuming a flat mass-ratio distribution.

\begin{figure}
    \centering
    \includegraphics[width=1.0\linewidth]{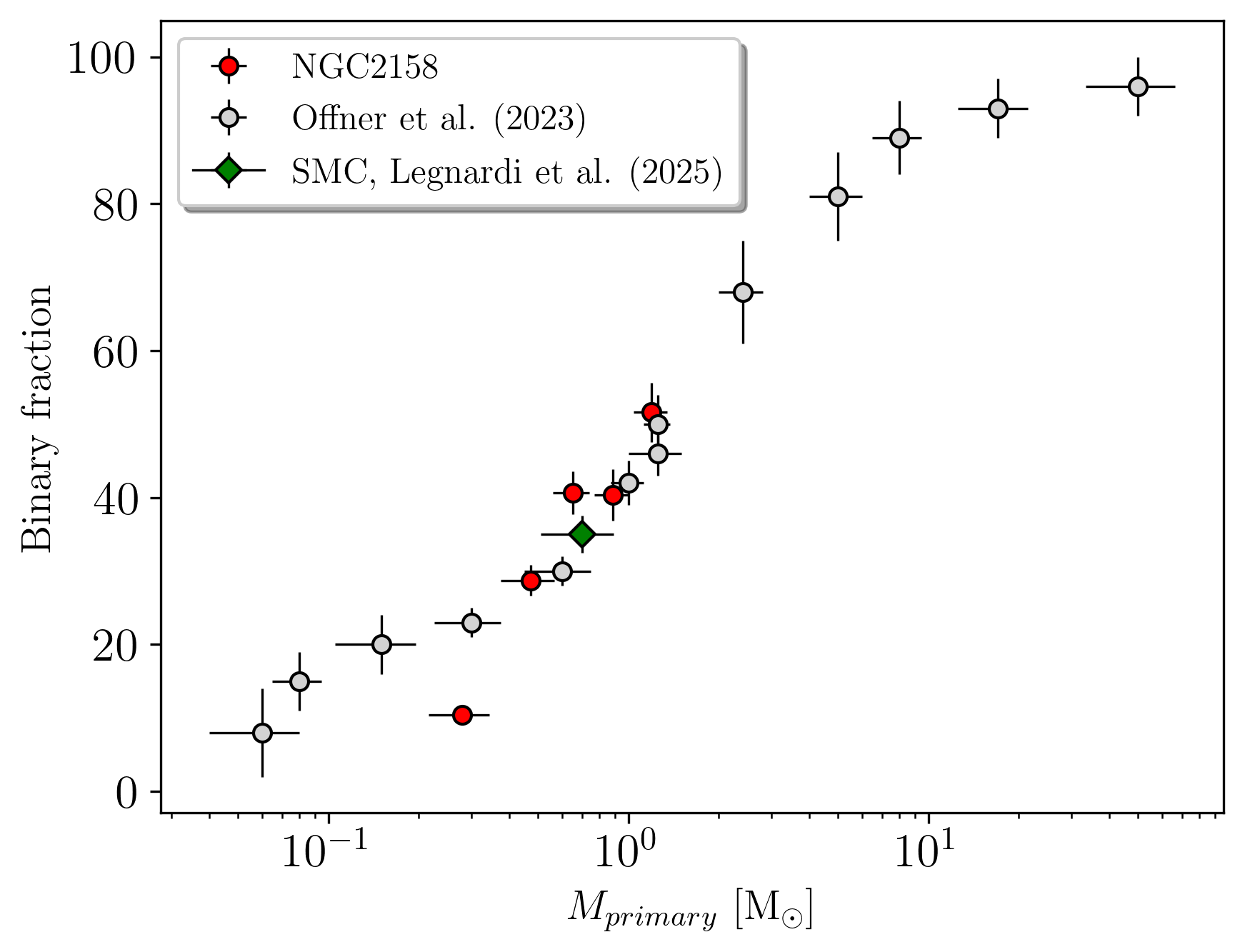}
    \caption{Total binary fraction as a function of primary-star mass from \cite{offner2023a} and \cite{legnardi2025a}, with the addition of five new data points (red dots) derived from our study of the open cluster NGC\,2158. The grey dots reproduce the literature compilation from \cite{offner2023a} for the Galaxy used in the original plot, and the green diamond represents the SMC binary fraction. }
    \label{fig:binaries_field}
\end{figure}

We find that more than 50\% of systems in the brightest magnitude bin are binaries, and that the binary fraction steadily decreases along the MS, reaching $\sim 20\%$ in the faintest bin analyzed. When comparing our results with those obtained in the literature for the Galactic field and for the Small Magellanic Cloud (SMC) in Fig.\,\ref{fig:binaries_field}, we find that they follow the same trend between total binary fraction and primary-star mass. This similarity suggests that binary formation and evolution in NGC\,2158 could be comparable to those observed in the low-density environments of the Milky Way and the SMC.

\begin{figure*}
    \centering
    \includegraphics[width=0.8\linewidth]{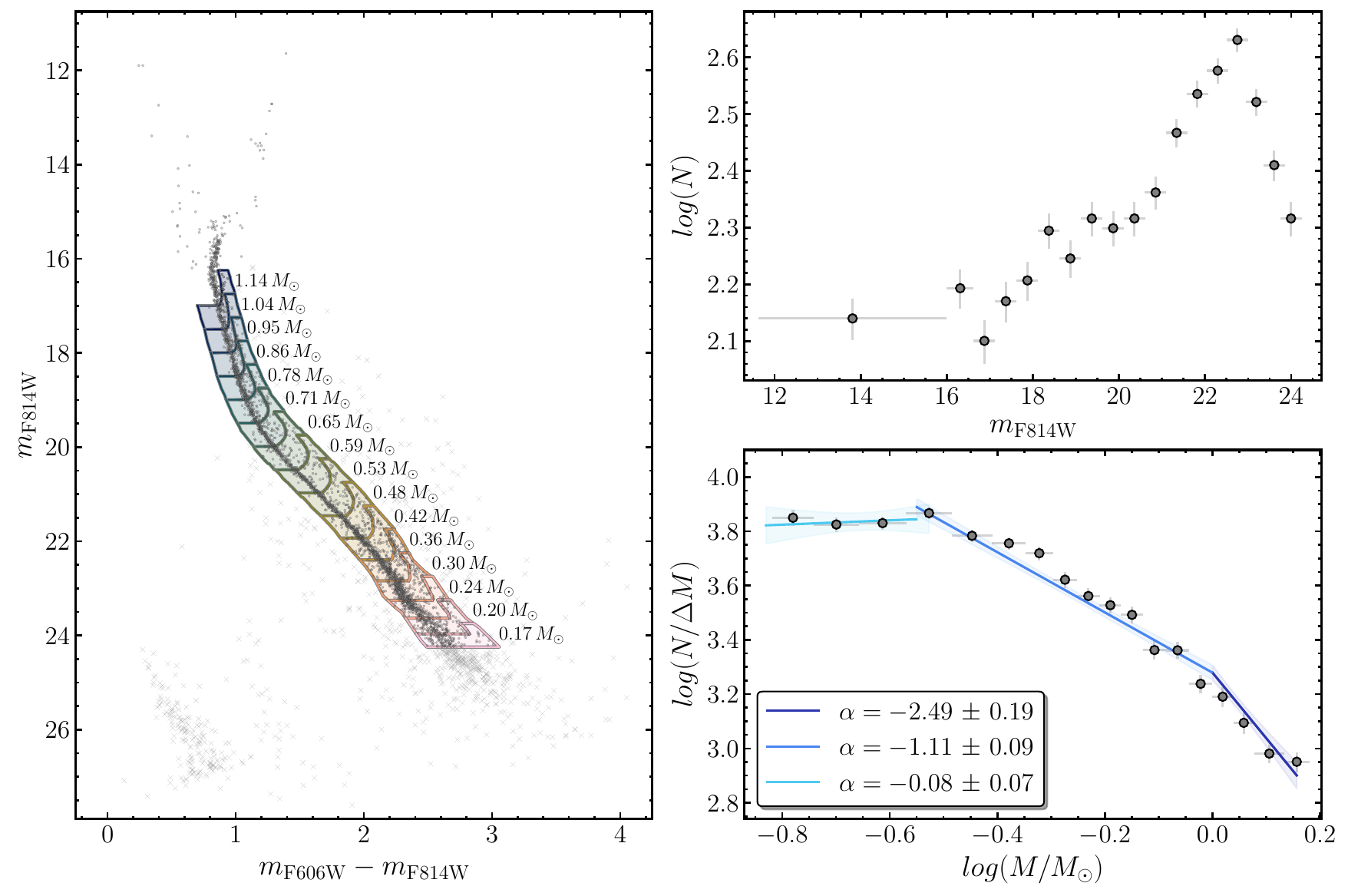}
    \caption{{\it Left}: Reproduction of the $m_{\rm F814W}$ versus $m_{\rm F606W} - m_{\rm F814W}$ CMD from Fig.\,\ref{fig:cmd}. The regions used to derive the luminosity and MFs are overplotted, with the corresponding average stellar masses indicated on the right side of each region. {\it Right}: Luminosity function (top) and MF (bottom) derived from the left-panel CMD. The straight lines show the best-fitting relations for stars in the mass intervals defined by \cite{cordoni2023a}. }
    \label{fig:MF}
\end{figure*}
\begin{table*}[h]
    \centering
        \caption{Fraction of binaries derived for stars in five luminosity intervals. For each interval we provide the stellar-mass range, the global fraction of binaries, and the fraction of binaries with q$>$0.5.}\label{tab:binaries}
    \begin{tabular}{c c l l }
        \hline
     $m_{F814W}$ range  & mass range  $[M_{\odot}]$ & $f_{bin}$ & $f_{bin}^{q>0.5}$ \\
         \hline      
17.0-18.6 & 0.83-1.12 & 0.52 $\pm$ 0.04 & 0.28 $\pm$ 0.02\\
18.6-20.2 & 0.61-0.83 & 0.40 $\pm$ 0.03 & 0.22 $\pm$ 0.02\\
20.2-21.8 & 0.43-0.61 & 0.41 $\pm$ 0.03 & 0.23 $\pm$ 0.02\\
21.8-23.4 & 0.25-0.43 & 0.29 $\pm$ 0.02 & 0.18 $\pm$ 0.01\\
23.4-25.0 & 0.14-0.25 & 0.10 $\pm$ 0.01 & 0.09 $\pm$ 0.01\\
        \hline
    \end{tabular}
    \label{tab:}
\end{table*}

\section{The mass function}\label{sec:MF}

To derive the stellar MF of NGC\,2158, we adopted the setup illustrated in Fig.\,\ref{fig:MF}, which is based on that used by \cite{legnardi2025a}. To analyze the MS, we defined seventeen regions in the CMD. Each region, constructed as described in Section\,\ref{sec:binaries}, includes single MS stars spanning a 0.25-mag-wide interval from $m_{\rm F814W}=17.0$ to 24.8, as well as binaries whose primary components fall within the same magnitude range. For binaries, the secondary-to-primary mass ratio ranges from 0 to 1.

Because of observational uncertainties, stars may scatter across adjacent magnitude–color bins, leading to contamination between neighboring regions in the CMD. Consequently, the total number of stars observed in each region (after completeness correction), $N_i$, can be expressed as
\begin{equation} \label{eq:LF}
N_i -N_{f,i} = \sum_{k} n_{k,i}, c_{k,i}
\end{equation}
where $N_{f,i}$ and $n_{k,i}$ are the number of field stars and the intrinsic number of stars in the $k$-th bin, and $c_{k,i}$ represents the fraction of those stars that, due to photometric errors, are scattered into the $i$-th bin. The contamination matrix $c_{k,i}$ was derived from synthetic CMDs constructed through AS experiments. 
These simulated CMDs adopt the same binary fractions measured in NGC\,2158, assuming a flat mass-ratio distribution.
 The intrinsic star counts $n_k$ were subsequently recovered by solving the linear system defined by Eq.\,\ref{eq:LF} \citep{milone2012b, legnardi2025a}.

The resulting luminosity function is shown in the top-right panel of Figure \ref{fig:MF}.  The error bars in each bin, which account for the Poisson noise of the observed star counts and the uncertainty in the completeness correction itself,  are estimated as 
$
\sigma_{Ni} 
= \sqrt{
\frac{N_{\rm i}}{C_{i}^2}
+
\frac{N_{\rm i}^2}{C_{i}^3}\,\frac{1-C_{i}}{N_{\rm AS, i}}
}.
$
where $C_{i}$ is the average completeness correction for the stars in the bin and $N_{\rm AS, i}$ the number of injected ASs.

The corresponding MF, derived by converting stellar magnitudes into masses using the relation between stellar mass and $m_{\rm F814W}$ magnitude from the best-fit isochrone, is presented in the bottom-right panel. A visual inspection of the MF reveals a clear change in slope around $1.0\,M_{\odot}$, consistent with the feature reported by \citet{cordoni2023a} in their analysis of Galactic open clusters based on Gaia DR3 data. Assuming a broken power-law with two segments joined at $1.0\,M_{\odot}$, we obtain slopes of $\alpha = -2.49 \pm 0.19$ and $\alpha = -1.11 \pm 0.09$ for stars less and more massive than this threshold, respectively and with masses larger than 0.3 M$_\odot$. At the lower end of the MS, for stellar masses below 0.3 M$_\odot$, outside the mass range analyzed by \citet{cordoni2023a}, the MF shows an additional flattening, characterized by a slope of $\alpha = -0.08 \pm 0.07$.

\begin{figure}
    \centering
    \includegraphics[width=1.0\linewidth]{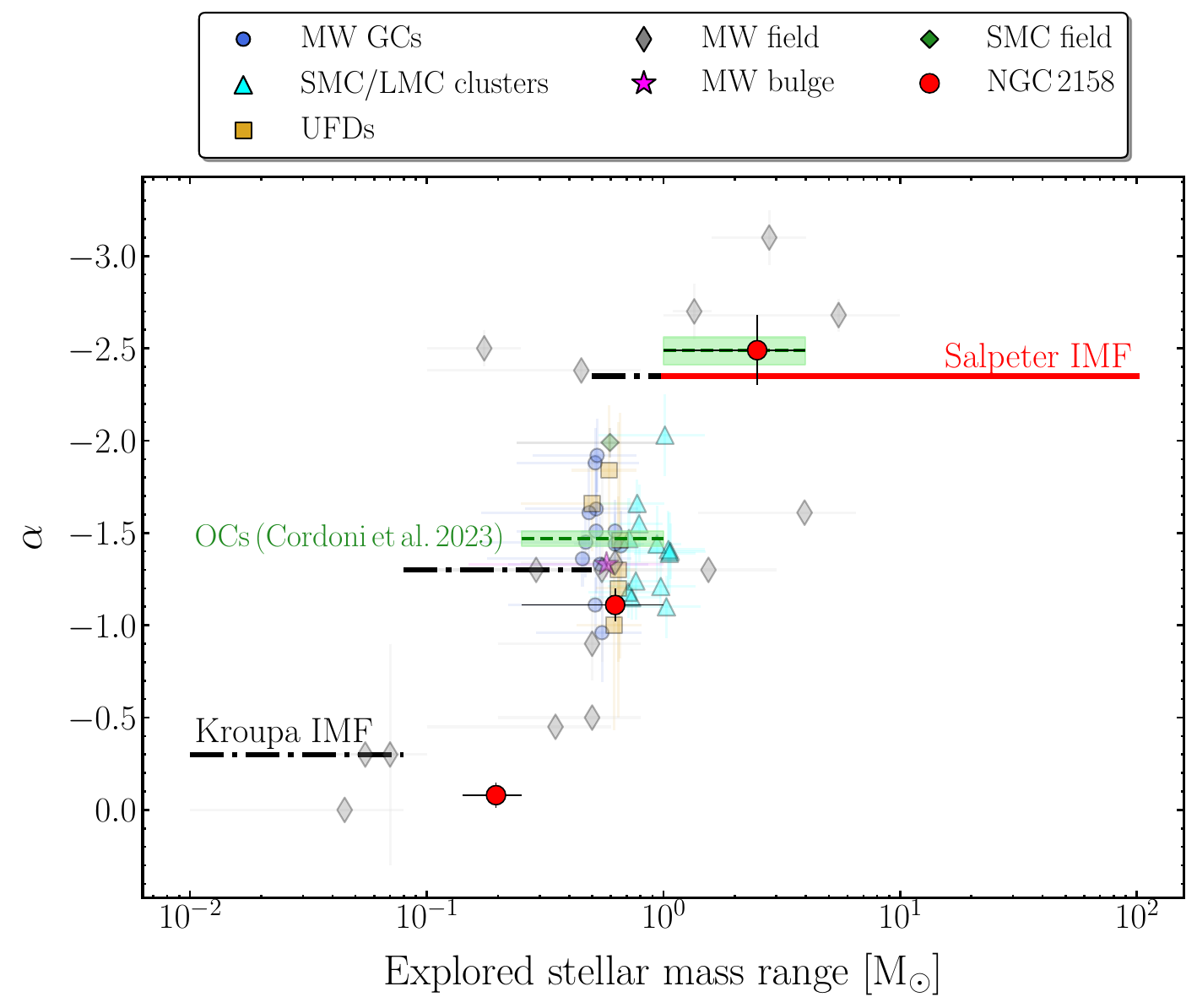}
    \caption{
 MF slope, $\alpha$, as a function of the explored stellar mass range, compiled from a variety of Galactic and extragalactic environments. Each point represents the best-fit power-law slope derived over a specific mass interval, with horizontal error bars indicating the width of the corresponding mass range. The green crosses indicate the measurements on NGC\,2158 from this work.  
 The solid red line marks the canonical \citet{salpeter1955a} IMF slope, while the dashed black lines represent the segmented IMF slopes from \citet{kroupa2001a}. The green dashed line corresponds to the slopes measured by \citet{cordoni2023a} for Galactic open clusters.
}
    \label{fig:MFs}
\end{figure}
If we instead impose a break at $M = 0.5\,M_{\odot}$, as in the canonical \citet{kroupa2001a} IMF, we derive slopes of $\alpha = -1.73 \pm 0.07$ for $M > 0.5\,M_{\odot}$ and $\alpha = -0.33 \pm 0.07$ for $M < 0.5\,M_{\odot}$. Both values are significantly shallower than the corresponding \citet{kroupa2001a} slopes ($\alpha = -2.3$ and $-1.3$), and we find no clear evidence for a distinct break at $M = 0.5\,M_{\odot}$. Overall, a representation with a break at $M = 1.0\,M_{\odot}$, similar to that adopted by \citet{cordoni2023a}, seems to provide a better description of the observed mass distribution than the Kroupa-like parametrization.

\begin{figure*}[htp!]
    \centering
    \includegraphics[width=0.85\linewidth]{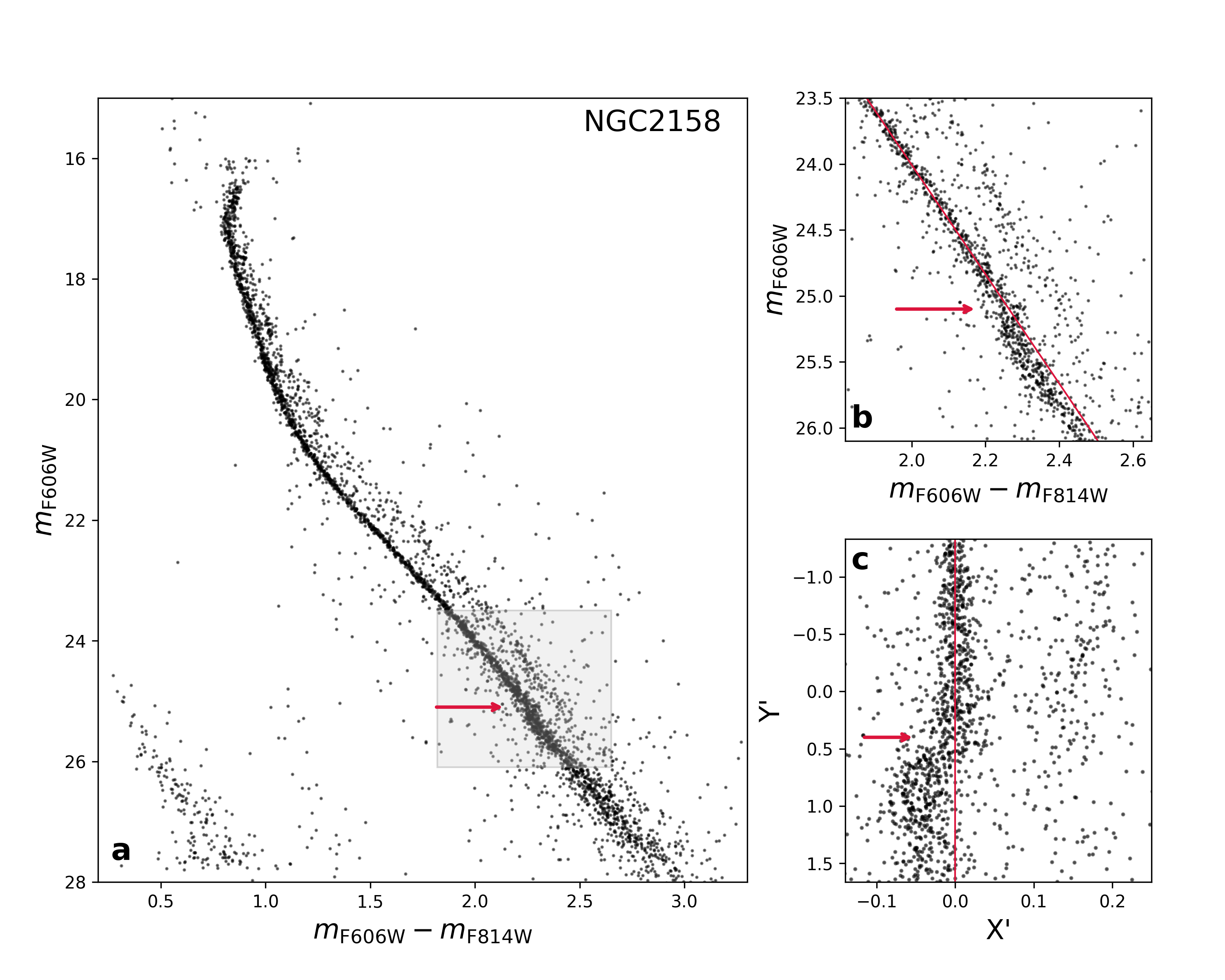}
    \caption{CMD of NGC\,2158 corrected for differential reddening, shown as $m_{\rm F606W}$ versus $m_{\rm F606W}-m_{\rm F814W}$ (panel a). Panel b presents a zoom-in of the low MS, corresponding to the gray rectangle in panel a. The red straight line marks the linear fit to the MS segment in the range $23.5<m_{\rm F606W}<24.75$. Panel c shows the CMD rotated such that this reference line becomes vertical. The arrows indicate the M-dwarf discontinuity.}
    \label{fig:cmdfeature}
\end{figure*}
\begin{figure}[htp!]
    \centering
    \includegraphics[width=1.\linewidth]{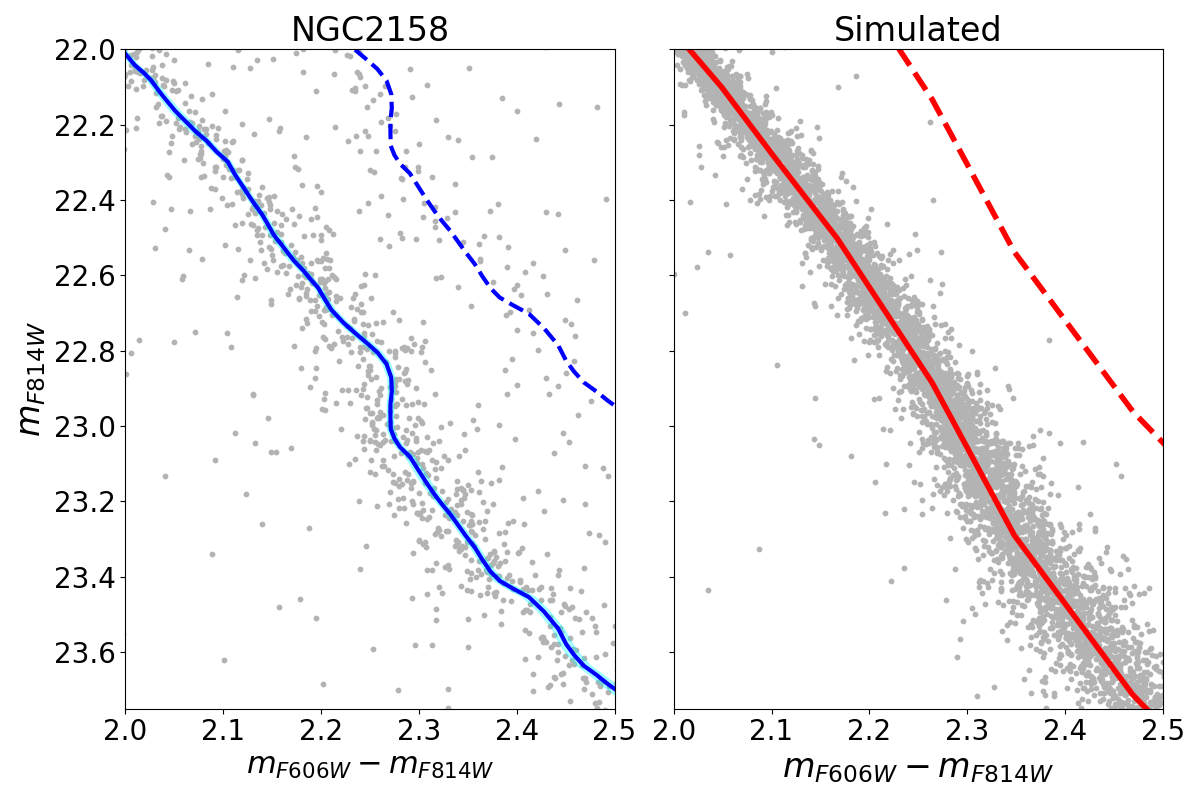}
    \caption{
    Zoomed-in view of the observed CMD of NGC\,2158 around the M-dwarf discontinuity. The blue solid line marks the MS fiducial, while the dashed line indicates the fiducial sequence for equal-mass binaries (left panel). The right panel shows the simulated CMD of MS stars; the red solid line represents the best-fit isochrone introduced in Fig.\,\ref{fig:cmd}, and the dashed line corresponds to the equal-mass MS–MS binary fiducial. }
    \label{fig:SIMU}
\end{figure}
The MF slope, $\alpha$, determined for NGC\,2158 is displayed in Fig.\,\ref{fig:MFs}, together with a comprehensive compilation of MF slopes measured for a wide variety of Galactic and extragalactic stellar systems. The comparison includes Galactic globular clusters (GCs; blue circles; \citealt{baumgardt2023a}), stellar clusters in the Small and Large Magellanic Clouds (SMC and LMC; crimson triangles; \citealt{baumgardt2023a}), open clusters (OCs; green dashed lines; \citealt{cordoni2023a}),\footnote{As in \cite{baumgardt2023a}, we consider only clusters whose lifetimes and relaxation times exceed their ages, ensuring that their present-day MFs still trace the initial ones.} ultra-faint dwarf galaxies (UFDs; gold squares; \citealt{geha2013a,gennaro2018a}), the Milky Way field (gray diamonds; \citealt{reid1999a,reid2002a,schroder2003a,kroupa2002a,allen2005a,metchev2008a,pinfield2008a,bochanski2010a,sollima2019a}), the Galactic bulge (magenta star; \citealt{zoccali2000a}), and OB associations (cyan thick crosses; \citealt{reyle2001a,schultheis2006a,vallenari2006a}).

 For completeness, we tested the impact of model-dependent systematics by repeating the LF–MF conversion using isochrones from the BaSTI \citep{pietrinferni2021a} and the Dartmouth \citep{dotter2008a} stellar evolution databases. In both cases we adopted the same age, metallicity, distance modulus, and reddening as for the reference model, which provide excellent match with the observed CMD. For the BaSTI isochrone, we adopted [$\alpha$/Fe]=0.2, as this slightly improves the fit to the observed data. The resulting MF slopes obtained from the BaSTI models for stars in the three analyzed mass bins are $\alpha = -2.49 \pm 0.17$, $ -0.97 \pm 0.07$, and $-0.22 \pm 0.07$, whereas the Dartmouth isochrones provide $\alpha = -2.49 \pm 0.20$, $ -1.01 \pm 0.09$, and $-0.04 \pm 0.09$
 These results are consistent within one-sigma and are therefore fully consistent with the reference values obtained with the models by \citet{marigo2017a}, demonstrating that our conclusions on the MFs are not sensitive to the adopted stellar models.

\section{A discontinuity along the M-dwarf sequence}\label{sec:gap}
A visual inspection of the CMD of NGC\,2158 shown in Figure\,\ref{fig:cmdfeature}a reveals a clear discontinuity in the MS slope around $m_{\rm F606W}=25.2$. At this magnitude, the MS abruptly turns toward bluer colors by approximately 0.05 mag for fainter magnitudes. A similar change in the slope appears to occur along the equal-mass binary sequence, about 0.75 mag brighter.
To better highlight this CMD feature, Figure\,\ref{fig:cmdfeature}b displays a zoomed-in view of the region surrounding the MS discontinuity. We fit a straight line to the MS segment above the discontinuity  ($23.5<m_{\rm F606W}<24.75$) and rotated the CMD around the point on this line corresponding to $m_{\rm F606W}=24.75$, so that the fit line becomes vertical. The resulting rotated diagram, plotted as $Y'$ versus $X'$ in Figure\,\ref{fig:cmdfeature}c, compresses the color spread of the selected MS stars, thus enhancing the visibility of subtle MS features that are less evident in the original CMD.
In the rotated diagram, the bulk of the MS, which exhibits an average pseudo-color of nearly zero around $Y'\sim0.4$, clearly shifts to negative $X'$ values below this threshold. The pseudo-color of MS stars appears to become closer to zero for pseudo-magnitudes fainter than $Y'\sim1.2$, thus hinting a further MS slope change for masses of about 0.22M$_{\odot}$.

To verify that the MS discontinuity is an intrinsic feature of the NGC\,2158 CMD, we compare in Fig.\,\ref{fig:SIMU} the observed CMD of faint MS stars with a simulated CMD. The latter was generated using ASs, as described in Sect.\,\ref{sec:data}, but in this case the ASs were placed along the best-fit isochrone from \citet{marigo2017a} (red continuous line in Fig.\,\ref{fig:SIMU}). Notably, neither the isochrone nor the simulated CMD reproduces the discontinuity observed in the real CMD around $m_{\rm F814W}=22.9$, as highlighted by the MS fiducial line in the left panel of Fig.\,\ref{fig:SIMU}. Thus, the MS discontinuity is not predicted by the best-fit isochrone, nor by isochrones from other databases \citep{dotter2008a, pietrinferni2021a}. Since the ASs were processed using the same reduction procedures and share the spatial and luminosity distribution as the real stars, this result demonstrates that the MS discontinuity is an intrinsic property of the cluster, rather than an artifact of the data reduction process.

We tentatively attribute this newly detected feature in the CMD, observed here for the first time in a star cluster, to the influence of the $^{3}$He-driven instabilities predicted by theoretical models of low-mass stars \citep{dantona1982a, vansaders2012a}. We will further discuss this interpretation in Section~\ref{sec:summary}.

\section{Summary and discussion}\label{sec:summary}

We used {\it HST} images of the Galactic open cluster NGC\,2158 to derive a deep CMD in the F606W and F814W bands from ACS/WFC observations. This diagram enables the identification of the cluster’s main evolutionary sequences, including the sparsely populated red clump, RGB, and sub-giant branch. The MS is clearly defined over a range of approximately ten magnitudes in F606W, extending down to stars with masses of about 0.1\,M$_{\odot}$. A prominent binary sequence is also visible, running redward of the MS. These deep {\it HST} data reveal the entire white-dwarf cooling sequence, which was the primary target of the {\it HST} programs under which this dataset was collected \citep{bedin2010a}. We take advantage of the exquisite photometry to constrain the MF, study the population of binaries, and investigate the MS morphology in the CMD. The main results can be summarized as follows: 

\begin{itemize}
    \item We derived the fraction of binary systems composed of two MS stars with mass ratios larger than $q = 0.5$. Our analysis covered a wide magnitude range, including single stars with masses between 0.14 and 1.12~M$_\odot$, as well as binary systems whose primary components span the same mass interval. We obtained a binary fraction of $f_{\rm bin}^{q>0.5} = 0.19 \pm 0.01$ and extrapolate a total fraction of binaries, $f_{\rm bin}=0.38 \pm 0.02$, consistent with the values observed in open clusters \citep[e.g.][]{cordoni2023a}. 
     We find no evidence for wide binaries with projected separations larger than $\sim$440 AU.
    We find that the total binary fraction decreases with stellar mass, from $0.52 \pm 0.04$ at $\sim$1.0~M$_\odot$ to $0.11 \pm 0.01$ at $\sim$0.2~M$_\odot$. This trend, together with the derived binary fractions, closely matches that observed in the Galactic field.
    \item The MF of NGC\,2158 is well represented by three distinct regimes: high-mass stars ($\alpha = -2.49 \pm 0.19$), low-mass stars ($\alpha = -1.11 \pm 0.09$), and very low-mass stars ($\alpha = -0.08 \pm 0.07$). The derived slopes and the turnover around $0.9\,M_{\odot}$ are consistent with those observed in the population of Galactic open clusters based on \textit{Gaia}~DR3 data \citep{cordoni2023a}. This agreement supports the conclusion by \citet{cordoni2023a} that open clusters exhibit remarkably similar MFs. In addition, we detect a pronounced deficit of stars at the lowest masses ($M \lesssim 0.3\,M_{\odot}$), a range not explored by \citet{cordoni2023a}, where we measure a nearly flat MF.

    \item We find that the MS fiducial line exhibits an abrupt change in slope at a stellar mass of approximately 0.3\,M$_{\odot}$. To our knowledge, such an MS discontinuity has not been reported in any other star cluster and we verified that is not predicted by current isochrone models \citep[][]{dotter2008a, marigo2017a, pietrinferni2021a}. To investigate the origin of this CMD feature, we examine theoretical studies and observational results for nearby Galactic stars.

\citet{dantona1982a} investigated convection in faint MS stars and showed that stars with masses of about $\sim$0.4 $M_{\odot}$ become fully convective, leading to total mixing. This mixing modifies the $^{3}$He concentration in the stellar core on short timescales, which produces a sudden change in nuclear luminosity and drives a thermal readjustment, accompanied by variations in radius, effective temperature, and luminosity.

Building on this line of investigation, \citet{vansaders2012a} predicted an instability in stars slightly more massive than the full-convection threshold \cite[see also][]{adronov2004a}. In their models, the buildup of central $^{3}$He causes the convective core to expand until it connects with the surface convection zone, triggering episodes of complete mixing, the so-called “convective kissing instability.” 

These events periodically alter the stellar structure, leading to modest ($\sim$ few percent) changes in radius and luminosity over Myr–Gyr timescales, but they eventually cease once the $^{3}$He abundance becomes sufficiently high for the star to remain fully convective for the rest of its MS lifetime. As a consequence, this instability, and any associated photometric discontinuity, is expected to be visible in young and intermediate-age stellar populations, but not in old clusters whose low-mass stars have already reached the fully convective phase. This is consistent with the fact that deep high-precision photometry of very low mass MS stars of several globular clusters and old open clusters did not show evidence of such feature in their CMDs \citep[e.g.\,][]{anderson2008b, bedin2008a, correnti2016a, ziliotto2023a, milone2025a}

It is plausible that the discontinuity in the NGC\,2158 CMD traces the same physical process predicted by models of $^{3}$He-driven instabilities in low-mass stars \citep{dantona1982a, vansaders2012a}. 

\end{itemize}

\begin{acknowledgements}
We thank the anonymous referee for the constructive suggestions that have improved the quality of the manuscript.
This research is based on observations made with the NASA/ESA Hubble Space Telescope obtained from the Space Telescope Science Institute, which is operated by the Association of Universities for Research in Astronomy, Inc., under NASA contract NAS 5–26555. These observations are associated with program 10500.
This work has been funded by the European Union – NextGenerationEU RRF M4C2 1.1 (PRIN 2022 2022MMEB9W: "Understanding the formation of globular clusters with their multiple stellar generations", CUP C53D23001200006), and from the European Union’s Horizon 2020 research and innovation programme under the Marie Skłodowska-Curie Grant Agreement No. 101034319 and from the European Union – NextGenerationEU (beneficiary: T. Ziliotto).
\end{acknowledgements}
\bibliographystyle{aa}
\bibliography{ms}

\begin{thebibliography}{66}
\expandafter\ifx\csname natexlab\endcsname\relax\def\natexlab#1{#1}\fi

\bibitem[{{Allen} {et~al.}(2005){Allen}, {Koerner}, {Reid}, \&
  {Trilling}}]{allen2005a}
{Allen}, P.~R., {Koerner}, D.~W., {Reid}, I.~N., \& {Trilling}, D.~E. 2005,
  \apj, 625, 385

\bibitem[{{Anderson}(2022)}]{anderson2022a}
{Anderson}, J. 2022, {One-Pass HST Photometry with hst1pass}, Instrument
  Science Report WFC3 2022-5, 55 pages

\bibitem[{{Anderson} \& {Bedin}(2010)}]{anderson2010a}
{Anderson}, J. \& {Bedin}, L.~R. 2010, \pasp, 122, 1035

\bibitem[{{Anderson} \& {King}(2000)}]{anderson2000a}
{Anderson}, J. \& {King}, I.~R. 2000, \pasp, 112, 1360

\bibitem[{{Anderson} \& {King}(2003)}]{anderson2003a}
{Anderson}, J. \& {King}, I.~R. 2003, \aj, 126, 772

\bibitem[{{Anderson} {et~al.}(2008{\natexlab{a}}){Anderson}, {King}, {Richer},
  {Fahlman}, {Hansen}, {Hurley}, {Kalirai}, {Rich}, \&
  {Stetson}}]{anderson2008b}
{Anderson}, J., {King}, I.~R., {Richer}, H.~B., {et~al.} 2008{\natexlab{a}},
  \aj, 135, 2114

\bibitem[{{Anderson} {et~al.}(2008{\natexlab{b}}){Anderson}, {Sarajedini},
  {Bedin}, {King}, {Piotto}, {Reid}, {Siegel}, {Majewski}, {Paust}, {Aparicio},
  {Milone}, {Chaboyer}, \& {Rosenberg}}]{anderson2008a}
{Anderson}, J., {Sarajedini}, A., {Bedin}, L.~R., {et~al.} 2008{\natexlab{b}},
  \aj, 135, 2055

\bibitem[{{Andronov} \& {Pinsonneault}(2004)}]{adronov2004a}
{Andronov}, N. \& {Pinsonneault}, M.~H. 2004, \apj, 614, 326

\bibitem[{{Baumgardt} {et~al.}(2023){Baumgardt}, {H{\'e}nault-Brunet},
  {Dickson}, \& {Sollima}}]{baumgardt2023a}
{Baumgardt}, H., {H{\'e}nault-Brunet}, V., {Dickson}, N., \& {Sollima}, A.
  2023, \mnras, 521, 3991

\bibitem[{{Bedin} {et~al.}(2010){Bedin}, {Salaris}, {King}, {Piotto},
  {Anderson}, \& {Cassisi}}]{bedin2010a}
{Bedin}, L.~R., {Salaris}, M., {King}, I.~R., {et~al.} 2010, \apjl, 708, L32

\bibitem[{{Bedin} {et~al.}(2008){Bedin}, {Salaris}, {Piotto}, {Cassisi},
  {Milone}, {Anderson}, \& {King}}]{bedin2008a}
{Bedin}, L.~R., {Salaris}, M., {Piotto}, G., {et~al.} 2008, \apjl, 679, L29

\bibitem[{{Bellazzini} {et~al.}(2002){Bellazzini}, {Fusi Pecci}, {Messineo},
  {Monaco}, \& {Rood}}]{bellazzini2002a}
{Bellazzini}, M., {Fusi Pecci}, F., {Messineo}, M., {Monaco}, L., \& {Rood},
  R.~T. 2002, \aj, 123, 1509

\bibitem[{{Bellini} {et~al.}(2017){Bellini}, {Anderson}, {Bedin}, {King}, {van
  der Marel}, {Piotto}, \& {Cool}}]{bellini2017a}
{Bellini}, A., {Anderson}, J., {Bedin}, L.~R., {et~al.} 2017, \apj, 842, 6

\bibitem[{{Bochanski} {et~al.}(2010){Bochanski}, {Hawley}, {Covey}, {West},
  {Reid}, {Golimowski}, \& {Ivezi{\'c}}}]{bochanski2010a}
{Bochanski}, J.~J., {Hawley}, S.~L., {Covey}, K.~R., {et~al.} 2010, \aj, 139,
  2679

\bibitem[{{B{\"o}hm-Vitense}(1970)}]{bohmvitense1970a}
{B{\"o}hm-Vitense}, E. 1970, \aap, 8, 283

\bibitem[{{Bohm-Vitense} \& {Canterna}(1974)}]{bohmvitense1974a}
{Bohm-Vitense}, E. \& {Canterna}, R. 1974, \apj, 194, 629

\bibitem[{{Bolte}(1992)}]{bolte1992a}
{Bolte}, M. 1992, \apjs, 82, 145

\bibitem[{Carraro {et~al.}(2002)Carraro, Girardi, \& Marigo}]{carraro2002a}
Carraro, G., Girardi, L., \& Marigo, P. 2002, Monthly Notices of the Royal
  Astronomical Society, 332, 705

\bibitem[{{Cool} \& {Bolton}(2002)}]{cool2002a}
{Cool}, A.~M. \& {Bolton}, A.~S. 2002, in Astronomical Society of the Pacific
  Conference Series, Vol. 263, Stellar Collisions, Mergers and their
  Consequences, ed. M.~M. {Shara}, 163

\bibitem[{Cordoni {et~al.}(2023)Cordoni, {Milone, Antonino P.}, {Marino, Anna
  F.}, {Vesperini, Enrico}, {Dondoglio, Emanuele}, {Legnardi, Maria Vittoria},
  {Mohandasan, Anjana}, {Carlos, Marilia}, {Lagioia, Edoardo P.}, {Jang,
  Sohee}, \& {Ziliotto, Tuila}}]{cordoni2023a}
Cordoni, G., {Milone, Antonino P.}, {Marino, Anna F.}, {et~al.} 2023, Astronomy
  \& Astrophysics, 672, A29

\bibitem[{{Correnti} {et~al.}(2016){Correnti}, {Gennaro}, {Kalirai}, {Brown},
  \& {Calamida}}]{correnti2016a}
{Correnti}, M., {Gennaro}, M., {Kalirai}, J.~S., {Brown}, T.~M., \& {Calamida},
  A. 2016, \apj, 823, 18

\bibitem[{{Dantona} \& {Mazzitelli}(1982)}]{dantona1982a}
{Dantona}, F. \& {Mazzitelli}, I. 1982, \apj, 260, 722

\bibitem[{{D'Antona} {et~al.}(2002){D'Antona}, {Montalb{\'a}n}, {Kupka}, \&
  {Heiter}}]{dantona2002a}
{D'Antona}, F., {Montalb{\'a}n}, J., {Kupka}, F., \& {Heiter}, U. 2002, \apjl,
  564, L93

\bibitem[{{de Bruijne} {et~al.}(2000){de Bruijne}, {Hoogerwerf}, \& {de
  Zeeuw}}]{debruijne2000a}
{de Bruijne}, J.~H.~J., {Hoogerwerf}, R., \& {de Zeeuw}, P.~T. 2000, \apjl,
  544, L65

\bibitem[{{de Bruijne} {et~al.}(2001){de Bruijne}, {Hoogerwerf}, \& {de
  Zeeuw}}]{debruijne2001a}
{de Bruijne}, J.~H.~J., {Hoogerwerf}, R., \& {de Zeeuw}, P.~T. 2001, \aap, 367,
  111

\bibitem[{{Dondoglio} {et~al.}(2022){Dondoglio}, {Milone}, {Renzini},
  {Vesperini}, {Lagioia}, {Marino}, {Bellini}, {Carlos}, {Cordoni}, {Jang},
  {Legnardi}, {Libralato}, {Mohandasan}, {D'Antona}, {Martorano}, {Muratore},
  \& {Tailo}}]{dondoglio2022a}
{Dondoglio}, E., {Milone}, A.~P., {Renzini}, A., {et~al.} 2022, \apj, 927, 207

\bibitem[{{Dotter} {et~al.}(2008){Dotter}, {Chaboyer}, {Jevremovi{\'c}},
  {Kostov}, {Baron}, \& {Ferguson}}]{dotter2008a}
{Dotter}, A., {Chaboyer}, B., {Jevremovi{\'c}}, D., {et~al.} 2008, \apjs, 178,
  89

\bibitem[{{Gaia Collaboration} {et~al.}(2021){Gaia Collaboration}, {Smart},
  {Sarro}, {Rybizki}, {Reyl{\'e}}, {Robin}, {Hambly}, {Abbas}, {Barstow}, {de
  Bruijne}, {Bucciarelli}, {Carrasco}, {Cooper}, {Hodgkin}, {Masana},
  {Michalik}, {Sahlmann}, {Sozzetti}, {Brown}, {Vallenari}, {Prusti},
  {Babusiaux}, {Biermann}, {Creevey}, {Evans}, {Eyer}, {Hutton}, {Jansen},
  {Jordi}, {Klioner}, {Lammers}, {Lindegren}, {Luri}, {Mignard}, {Panem},
  {Pourbaix}, {Randich}, {Sartoretti}, {Soubiran}, {Walton}, {Arenou},
  {Bailer-Jones}, {Bastian}, {Cropper}, {Drimmel}, {Katz}, {Lattanzi}, {van
  Leeuwen}, {Bakker}, {Casta{\~n}eda}, {De Angeli}, {Ducourant}, {Fabricius},
  {Fouesneau}, {Fr{\'e}mat}, {Guerra}, {Guerrier}, {Guiraud}, {Jean-Antoine
  Piccolo}, {Messineo}, {Mowlavi}, {Nicolas}, {Nienartowicz}, {Pailler},
  {Panuzzo}, {Riclet}, {Roux}, {Seabroke}, {Sordo}, {Tanga}, {Th{\'e}venin},
  {Gracia-Abril}, {Portell}, {Teyssier}, {Altmann}, {Andrae}, {Bellas-Velidis},
  {Benson}, {Berthier}, {Blomme}, {Brugaletta}, {Burgess}, {Busso}, {Carry},
  {Cellino}, {Cheek}, {Clementini}, {Damerdji}, {Davidson}, {Delchambre},
  {Dell'Oro}, {Fern{\'a}ndez-Hern{\'a}ndez}, {Galluccio}, {Garc{\'\i}a-Lario},
  {Garcia-Reinaldos}, {Gonz{\'a}lez-N{\'u}{\~n}ez}, {Gosset}, {Haigron},
  {Halbwachs}, {Harrison}, {Hatzidimitriou}, {Heiter}, {Hern{\'a}ndez},
  {Hestroffer}, {Holl}, {Jan{\ss}en}, {Jevardat de Fombelle}, {Jordan},
  {Krone-Martins}, {Lanzafame}, {L{\"o}ffler}, {Lorca}, {Manteiga}, {Marchal},
  {Marrese}, {Moitinho}, {Mora}, {Muinonen}, {Osborne}, {Pancino}, {Pauwels},
  {Recio-Blanco}, {Richards}, {Riello}, {Rimoldini}, {Roegiers}, {Siopis},
  {Smith}, {Ulla}, {Utrilla}, {van Leeuwen}, {van Reeven}, {Abreu Aramburu},
  {Accart}, {Aerts}, {Aguado}, {Ajaj}, {Altavilla}, {{\'A}lvarez}, {{\'A}lvarez
  Cid-Fuentes}, {Alves}, {Anderson}, {Anglada Varela}, {Antoja}, {Audard},
  {Baines}, {Baker}, {Balaguer-N{\'u}{\~n}ez}, {Balbinot}, {Balog}, {Barache},
  {Barbato}, {Barros}, {Bartolom{\'e}}, {Bassilana}, {Bauchet},
  {Baudesson-Stella}, {Becciani}, {Bellazzini}, {Bernet}, {Bertone}, {Bianchi},
  {Blanco-Cuaresma}, {Boch}, {Bombrun}, {Bossini}, {Bouquillon}, {Bragaglia},
  {Bramante}, {Breedt}, {Bressan}, {Brouillet}, {Burlacu}, {Busonero},
  {Butkevich}, {Buzzi}, {Caffau}, {Cancelliere}, {C{\'a}novas},
  {Cantat-Gaudin}, {Carballo}, {Carlucci}, {Carnerero}, {Casamiquela},
  {Castellani}, {Castro-Ginard}, {Castro Sampol}, {Chaoul}, {Charlot},
  {Chemin}, {Chiavassa}, {Cioni}, {Comoretto}, {Cornez}, {Cowell}, {Crifo},
  {Crosta}, {Crowley}, {Dafonte}, \& {Dapergolas}}]{gaia2021a}
{Gaia Collaboration}, {Smart}, R.~L., {Sarro}, L.~M., {et~al.} 2021, \aap, 649,
  A6

\bibitem[{{Geha} {et~al.}(2013){Geha}, {Brown}, {Tumlinson}, {Kalirai},
  {Simon}, {Kirby}, {VandenBerg}, {Mu{\~n}oz}, {Avila}, {Guhathakurta}, \&
  {Ferguson}}]{geha2013a}
{Geha}, M., {Brown}, T.~M., {Tumlinson}, J., {et~al.} 2013, \apj, 771, 29

\bibitem[{{Gennaro} {et~al.}(2018){Gennaro}, {Tchernyshyov}, {Brown}, {Geha},
  {Avila}, {Guhathakurta}, {Kalirai}, {Kirby}, {Renzini}, {Simon}, {Tumlinson},
  \& {Vargas}}]{gennaro2018a}
{Gennaro}, M., {Tchernyshyov}, K., {Brown}, T.~M., {et~al.} 2018, \apj, 855, 20

\bibitem[{{Girardi} {et~al.}(2005){Girardi}, {Groenewegen}, {Hatziminaoglou},
  \& {da Costa}}]{girardi2005a}
{Girardi}, L., {Groenewegen}, M.~A.~T., {Hatziminaoglou}, E., \& {da Costa}, L.
  2005, \aap, 436, 895

\bibitem[{{Jao} {et~al.}(2018){Jao}, {Henry}, {Gies}, \& {Hambly}}]{jao2018a}
{Jao}, W.-C., {Henry}, T.~J., {Gies}, D.~R., \& {Hambly}, N.~C. 2018, \apjl,
  861, L11

\bibitem[{{Kroupa}(2001)}]{kroupa2001a}
{Kroupa}, P. 2001, \mnras, 322, 231

\bibitem[{{Kroupa} \& {Boily}(2002)}]{kroupa2002a}
{Kroupa}, P. \& {Boily}, C.~M. 2002, \mnras, 336, 1188

\bibitem[{{Legnardi} {et~al.}(2023){Legnardi}, {Milone}, {Cordoni}, {Lagioia},
  {Dondoglio}, {Marino}, {Jang}, {Mohandasan}, \& {Ziliotto}}]{legnardi2023a}
{Legnardi}, M.~V., {Milone}, A.~P., {Cordoni}, G., {et~al.} 2023, \mnras, 522,
  367

\bibitem[{{Legnardi} {et~al.}(2025){Legnardi}, {Muratore}, {Milone}, {Cordoni},
  {Ziliotto}, {Dondoglio}, {Marino}, {Mastrobuono-Battisti}, {Bortolan},
  {Lagioia}, \& {Tailo}}]{legnardi2025a}
{Legnardi}, M.~V., {Muratore}, F., {Milone}, A.~P., {et~al.} 2025, \aap, 702,
  A180

\bibitem[{{Marigo} {et~al.}(2017){Marigo}, {Girardi}, {Bressan}, {Rosenfield},
  {Aringer}, {Chen}, {Dussin}, {Nanni}, {Pastorelli}, {Rodrigues}, {Trabucchi},
  {Bladh}, {Dalcanton}, {Groenewegen}, {Montalb{\'a}n}, \&
  {Wood}}]{marigo2017a}
{Marigo}, P., {Girardi}, L., {Bressan}, A., {et~al.} 2017, \apj, 835, 77

\bibitem[{{Marino} {et~al.}(2024){Marino}, {Milone}, {Legnardi}, {Renzini},
  {Dondoglio}, {Cavecchi}, {Cordoni}, {Dotter}, {Lagioia}, {Ziliotto},
  {Bernizzoni}, {Bortolan}, {Carlos}, {Jang}, {Mohandasan}, {Muratore}, \&
  {Tailo}}]{marino2024a}
{Marino}, A.~F., {Milone}, A.~P., {Legnardi}, M.~V., {et~al.} 2024, \apj, 965,
  189

\bibitem[{{Metchev} {et~al.}(2008){Metchev}, {Kirkpatrick}, {Berriman}, \&
  {Looper}}]{metchev2008a}
{Metchev}, S.~A., {Kirkpatrick}, J.~D., {Berriman}, G.~B., \& {Looper}, D.
  2008, \apj, 676, 1281

\bibitem[{{Milone} {et~al.}(2025){Milone}, {Cordoni}, {Marino}, {Altomonte},
  {Dondoglio}, {Legnardi}, {Bortolan}, {Lionetto}, {Marchuk}, {Muratore}, \&
  {Ziliotto}}]{milone2025a}
{Milone}, A.~P., {Cordoni}, G., {Marino}, A.~F., {et~al.} 2025, \aap, 696, A221

\bibitem[{{Milone} {et~al.}(2023){Milone}, {Cordoni}, {Marino}, {D'Antona},
  {Bellini}, {Di Criscienzo}, {Dondoglio}, {Lagioia}, {Langer}, {Legnardi},
  {Libralato}, {Baumgardt}, {Bettinelli}, {Cavecchi}, {de Grijs}, {Deng},
  {Hastings}, {Li}, {Mohandasan}, {Renzini}, {Vesperini}, {Wang}, {Ziliotto},
  {Carlos}, {Costa}, {Dell'Agli}, {Di Stefano}, {Jang}, {Martorano}, {Simioni},
  {Tailo}, \& {Ventura}}]{milone2023a}
{Milone}, A.~P., {Cordoni}, G., {Marino}, A.~F., {et~al.} 2023, \aap, 672, A161

\bibitem[{{Milone} {et~al.}(2016){Milone}, {Marino}, {Bedin}, {Dotter},
  {Jerjen}, {Kim}, {Nardiello}, {Piotto}, \& {Cong}}]{milone2016a}
{Milone}, A.~P., {Marino}, A.~F., {Bedin}, L.~R., {et~al.} 2016, \mnras, 455,
  3009

\bibitem[{{Milone} {et~al.}(2012{\natexlab{a}}){Milone}, {Piotto}, {Bedin},
  {Aparicio}, {Anderson}, {Sarajedini}, {Marino}, {Moretti}, {Davies},
  {Chaboyer}, {Dotter}, {Hempel}, {Mar{\'\i}n-Franch}, {Majewski}, {Paust},
  {Reid}, {Rosenberg}, \& {Siegel}}]{milone2012a}
{Milone}, A.~P., {Piotto}, G., {Bedin}, L.~R., {et~al.} 2012{\natexlab{a}},
  \aap, 540, A16

\bibitem[{{Milone} {et~al.}(2012{\natexlab{b}}){Milone}, {Piotto}, {Bedin},
  {Cassisi}, {Anderson}, {Marino}, {Pietrinferni}, \& {Aparicio}}]{milone2012b}
{Milone}, A.~P., {Piotto}, G., {Bedin}, L.~R., {et~al.} 2012{\natexlab{b}},
  \aap, 537, A77

\bibitem[{{Mohandasan} {et~al.}(2024){Mohandasan}, {Milone}, {Cordoni},
  {Dondoglio}, {Lagioia}, {Legnardi}, {Ziliotto}, {Jang}, {Marino}, \&
  {Carlos}}]{mohandasan2024a}
{Mohandasan}, A., {Milone}, A.~P., {Cordoni}, G., {et~al.} 2024, \aap, 681, A42

\bibitem[{{Muratore} {et~al.}(2025){Muratore}, {Legnardi}, {Milone},
  {Mastrobuono-Battisti}, {Cordoni}, {Gorza}, {Lagioia}, {Bortolan},
  {Dondoglio}, {Marino}, \& {Ziliotto}}]{muratore2025a}
{Muratore}, F., {Legnardi}, M.~V., {Milone}, A.~P., {et~al.} 2025, arXiv
  e-prints, arXiv:2512.01547

\bibitem[{{Offner} {et~al.}(2023){Offner}, {Moe}, {Kratter}, {Sadavoy},
  {Jensen}, \& {Tobin}}]{offner2023a}
{Offner}, S.~S.~R., {Moe}, M., {Kratter}, K.~M., {et~al.} 2023, in Astronomical
  Society of the Pacific Conference Series, Vol. 534, Protostars and Planets
  VII, ed. S.~{Inutsuka}, Y.~{Aikawa}, T.~{Muto}, K.~{Tomida}, \& M.~{Tamura},
  275

\bibitem[{{Pietrinferni} {et~al.}(2021){Pietrinferni}, {Hidalgo}, {Cassisi},
  {Salaris}, {Savino}, {Mucciarelli}, {Verma}, {Silva Aguirre}, {Aparicio}, \&
  {Ferguson}}]{pietrinferni2021a}
{Pietrinferni}, A., {Hidalgo}, S., {Cassisi}, S., {et~al.} 2021, \apj, 908, 102

\bibitem[{{Pinfield} {et~al.}(2008){Pinfield}, {Burningham}, {Tamura},
  {Leggett}, {Lodieu}, {Lucas}, {Mortlock}, {Warren}, {Homeier}, {Ishii},
  {Deacon}, {McMahon}, {Hewett}, {Osori}, {Martin}, {Jones}, {Venemans},
  {Day-Jones}, {Dobbie}, {Folkes}, {Dye}, {Allard}, {Baraffe}, {Barrado Y
  Navascu{\'e}s}, {Casewell}, {Chiu}, {Chabrier}, {Clarke}, {Hodgkin},
  {Magazz{\`u}}, {McCaughrean}, {Nakajima}, {Pavlenko}, \&
  {Tinney}}]{pinfield2008a}
{Pinfield}, D.~J., {Burningham}, B., {Tamura}, M., {et~al.} 2008, \mnras, 390,
  304

\bibitem[{{Piotto} {et~al.}(2012){Piotto}, {Milone}, {Anderson}, {Bedin},
  {Bellini}, {Cassisi}, {Marino}, {Aparicio}, \& {Nascimbeni}}]{piotto2012a}
{Piotto}, G., {Milone}, A.~P., {Anderson}, J., {et~al.} 2012, \apj, 760, 39

\bibitem[{{Raghavan} {et~al.}(2010){Raghavan}, {McAlister}, {Henry}, {Latham},
  {Marcy}, {Mason}, {Gies}, {White}, \& {ten Brummelaar}}]{raghavan2010a}
{Raghavan}, D., {McAlister}, H.~A., {Henry}, T.~J., {et~al.} 2010, \apjs, 190,
  1

\bibitem[{{Reid} {et~al.}(2002){Reid}, {Gizis}, \& {Hawley}}]{reid2002a}
{Reid}, I.~N., {Gizis}, J.~E., \& {Hawley}, S.~L. 2002, \aj, 124, 2721

\bibitem[{{Reid} {et~al.}(1999){Reid}, {Kirkpatrick}, {Liebert}, {Burrows},
  {Gizis}, {Burgasser}, {Dahn}, {Monet}, {Cutri}, {Beichman}, \&
  {Skrutskie}}]{reid1999a}
{Reid}, I.~N., {Kirkpatrick}, J.~D., {Liebert}, J., {et~al.} 1999, \apj, 521,
  613

\bibitem[{{Reyl{\'e}} \& {Robin}(2001)}]{reyle2001a}
{Reyl{\'e}}, C. \& {Robin}, A.~C. 2001, \aap, 373, 886

\bibitem[{{Richer} {et~al.}(2006){Richer}, {Anderson}, {Brewer}, {Davis},
  {Fahlman}, {Hansen}, {Hurley}, {Kalirai}, {King}, {Reitzel}, {Rich}, {Shara},
  \& {Stetson}}]{richer2006a}
{Richer}, H.~B., {Anderson}, J., {Brewer}, J., {et~al.} 2006, Science, 313, 936

\bibitem[{{Romani} \& {Weinberg}(1991)}]{romani1991a}
{Romani}, R.~W. \& {Weinberg}, M.~D. 1991, \apj, 372, 487

\bibitem[{{Sabbi} {et~al.}(2016){Sabbi}, {Lennon}, {Anderson}, {Cignoni}, {van
  der Marel}, {Zaritsky}, {De Marchi}, {Panagia}, {Gouliermis}, {Grebel},
  {Gallagher}, {Smith}, {Sana}, {Aloisi}, {Tosi}, {Evans}, {Arab}, {Boyer}, {de
  Mink}, {Gordon}, {Koekemoer}, {Larsen}, {Ryon}, \& {Zeidler}}]{sabbi2016a}
{Sabbi}, E., {Lennon}, D.~J., {Anderson}, J., {et~al.} 2016, \apjs, 222, 11

\bibitem[{{Salpeter}(1955)}]{salpeter1955a}
{Salpeter}, E.~E. 1955, \apj, 121, 161

\bibitem[{{Schr{\"o}der} \& {Pagel}(2003)}]{schroder2003a}
{Schr{\"o}der}, K.-P. \& {Pagel}, B.~E.~J. 2003, \mnras, 343, 1231

\bibitem[{{Schultheis} {et~al.}(2006){Schultheis}, {Robin}, {Reyl{\'e}},
  {McCracken}, {Bertin}, {Mellier}, \& {Le F{\`e}vre}}]{schultheis2006a}
{Schultheis}, M., {Robin}, A.~C., {Reyl{\'e}}, C., {et~al.} 2006, \aap, 447,
  185

\bibitem[{{Sollima}(2019)}]{sollima2019a}
{Sollima}, A. 2019, \mnras, 489, 2377

\bibitem[{{Sollima} {et~al.}(2007){Sollima}, {Beccari}, {Ferraro}, {Fusi
  Pecci}, \& {Sarajedini}}]{sollima2007a}
{Sollima}, A., {Beccari}, G., {Ferraro}, F.~R., {Fusi Pecci}, F., \&
  {Sarajedini}, A. 2007, \mnras, 380, 781

\bibitem[{{Vallenari} {et~al.}(2006){Vallenari}, {Pasetto}, {Bertelli},
  {Chiosi}, {Spagna}, \& {Lattanzi}}]{vallenari2006a}
{Vallenari}, A., {Pasetto}, S., {Bertelli}, G., {et~al.} 2006, \aap, 451, 125

\bibitem[{{van Saders} \& {Pinsonneault}(2012)}]{vansaders2012a}
{van Saders}, J.~L. \& {Pinsonneault}, M.~H. 2012, \apj, 751, 98

\bibitem[{{Ziliotto} {et~al.}(2023){Ziliotto}, {Milone}, {Marino}, {Dotter},
  {Renzini}, {Vesperini}, {Karakas}, {Cordoni}, {Dondoglio}, {Legnardi},
  {Lagioia}, {Mohandasan}, \& {Baimukhametova}}]{ziliotto2023a}
{Ziliotto}, T., {Milone}, A., {Marino}, A.~F., {et~al.} 2023, \apj, 953, 62

\bibitem[{{Zoccali} {et~al.}(2000){Zoccali}, {Cassisi}, {Frogel}, {Gould},
  {Ortolani}, {Renzini}, {Rich}, \& {Stephens}}]{zoccali2000a}
{Zoccali}, M., {Cassisi}, S., {Frogel}, J.~A., {et~al.} 2000, \apj, 530, 418

\end{thebibliography}

\newpage
\appendix
\section{Proper motions}
\label{app:pms}

 The availability of HST images obtained at multiple epochs enables the measurement of stellar proper motions. We use these proper motions to distinguish the bulk of cluster members from field stars and to determine the binary fraction for the sample of NGC\,2158 cluster stars. At odds with what we have done in Sect.\ref{sec:binaries}, where we used the Trilegal Galactic model \citep{girardi2005a} to estimate the numbers of field stars in the observed CMD, the fraction of binaries inferred in this section rely on proper-motion selected field stars and cluster members.

\begin{figure}
    \centering
    \includegraphics[width=1.25\linewidth]{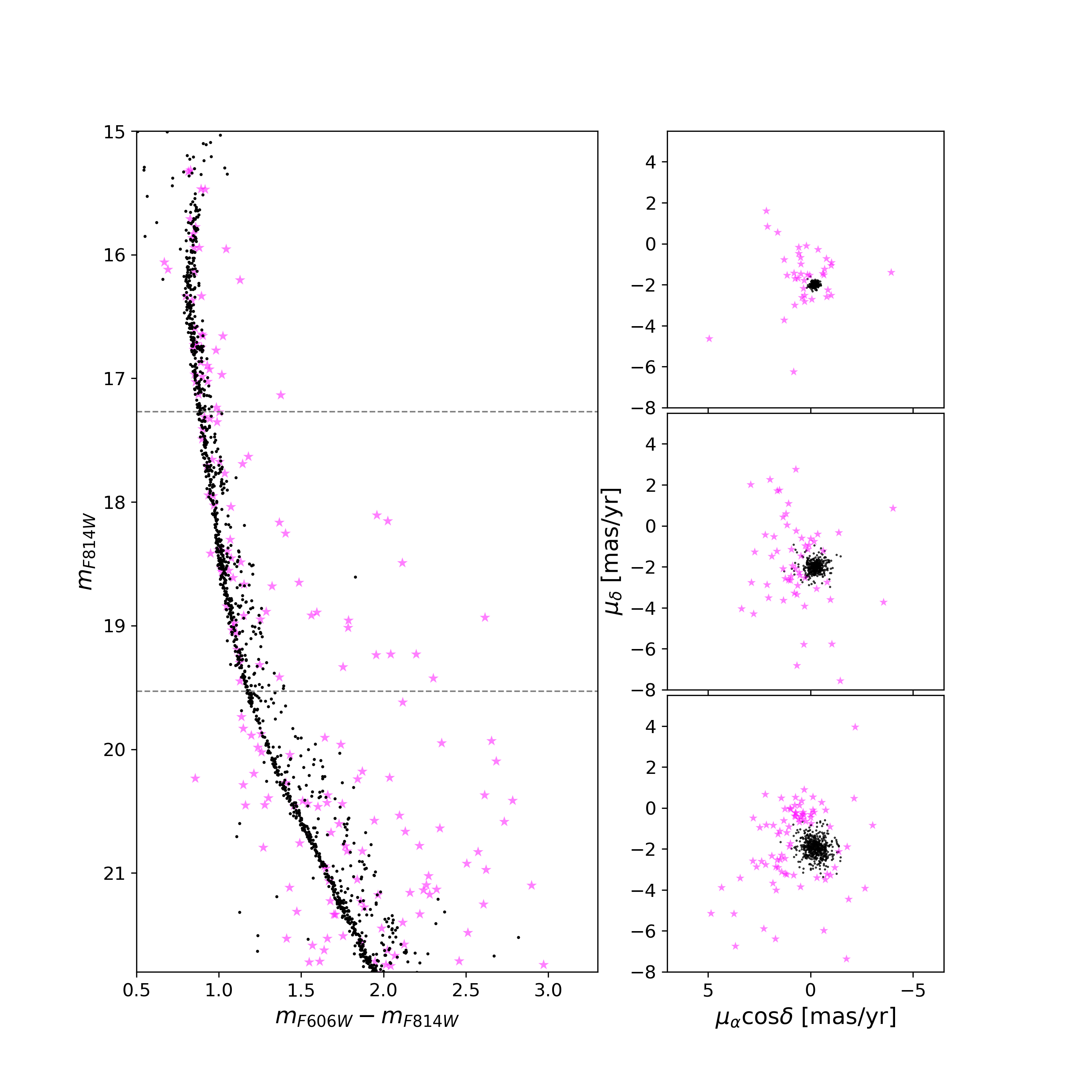}
\caption{Left panel: Reproduction of the $m_{\rm F814W}$ versus
$(m_{\rm F606W}-m_{\rm F814W})$ CMD of
NGC\,2158 for stars brighter than $m_{\rm F814W}=21.8$ with measured
proper motions. Right panels: Proper-motion diagrams for stars in the
three $m_{\rm F814W}$ magnitude intervals indicated by the horizontal
dotted lines in the CMD. Probable cluster members are shown as black
dots, while field stars are marked by magenta starred symbols.}
    \label{fig:pms}
\end{figure}

To derive stellar proper motions we followed the approach described by \citet[][Sect.\,5]{milone2023a} \citep[see also][]{anderson2003a, piotto2012a}. First, we considered astrometric catalogs for each epoch by using stellar positions in all available {\it HST} images collected in 2005 and 2006 and the stellar position provided by the Gaia DR3 catalog \citep{gaia2021a}. We transformed positions into a common master reference frame using linear six-parameter transformations calibrated on bright, well-measured stars. For each star, its transformed positions as a function of observational epoch were fitted with a weighted least-squares line, the slope of which provides the relative proper motion. The relative proper motions have been converted into absolute ones by comparing them with the absolute stellar proper motions from Gaia DR3.  The probable cluster members were defined iteratively as stars with motions consistent with the cluster bulk motion \citep[see][for details]{milone2023a}.
The results are illustrated in Fig.\,\ref{fig:pms}, where we show the  $m_{\rm F814W}$ versus $(m_{\rm F606W}-m_{\rm F814W})$ CMD (left panel) together with the proper-motion diagram of stars in three F814W magnitude intervals (right panels). We used black dots and magenta starred symbols to represent probable cluster members and field stars, respectively.
Noticeably, we limited the analysis to stars with $m_{\rm F814W}<21.8$. Indeed, due to the short time baseline of about one year of the HST images, it is not possible to derive precise proper motions for faint stars.

Based on the CMDs of proper-motion--selected field stars and cluster members, we estimated the fraction of binaries with mass ratios (q$>$0.5) in three F814W magnitude intervals, (17.0-18.6), (18.6-20.2), and (20.2-21.8), using the procedure described in Sect.\,\ref{sec:binaries} (see also Sect.~4 of \citealt{milone2012a}). We obtained binary fractions of (0.26$\pm$0.03), (0.24$\pm$0.03), and (0.21$\pm$0.02), respectively. These values agree within 1$\sigma$ with those reported in Table~\ref{tab:binaries}, which were derived using simulated field stars from \texttt{TRILEGAL}. This agreement confirms that our results are not significantly affected by the choice of the adopted field-star sample.

\label{lastpage}
\end{document}